\documentclass[aps,prb,graphicx,preprint]{revtex4-2}

\usepackage{hyperref}
\usepackage{amsmath}
\usepackage{array}
\usepackage{dcolumn}
\usepackage{graphicx}
\usepackage{grffile}
\usepackage{siunitx}
\usepackage{mw}
\usepackage{bm}
\usepackage{xcolor}
\usepackage{multirow}
\begin{document}

\title{Analysis of \textit{ab-initio} total energies obtained by different DFT implementations}
\author{Vishnu Raghuraman}
\affiliation{Department of Physics, Carnegie Mellon University, Pittsburgh, PA 15213}

\author{Yang Wang}
\affiliation{Pittsburgh Supercomputing Center, Carnegie Mellon University, Pittsburgh, PA 15213}

\author{Michael Widom}
\affiliation{Department of Physics, Carnegie Mellon University, Pittsburgh, PA 15213}

\begin{abstract}
Ab-initio crystal structure prediction depends on accurate calculation of the energies of competing structures. Many DFT codes are available that utilize different approaches to solve the Kohn-Sham equation.  We evaluate the consistency of three software packages (WIEN2k, VASP and MuST) that utilize three different methods (FL-APW, plane-wave pseudopotential and the KKR-Green's Function methods) by comparing the relative  total energies obtained for a set of BCC and FCC binary metallic alloys. We focus on the impact of choices such as muffin-tin {\em{vs.}} full-potential, angular momentum cutoff and other important KKR parameters. Different alloy systems prove more or less sensitive to these choices, and we explain the differences through study of the angular variation of their potentials. Our results can provide guidance in the application of KKR as a total energy method for structure prediction.
\end{abstract}
\maketitle

\section{Introduction}
Density functional theory-based first principles calculations are a powerful tool to uncover useful and interesting functional properties, guide alloy design and explain experimentally observed behavior in a variety of condensed matter systems\cite{dft,quantumchem,miracle,yeh}. This approach is highly popular due to (among other reasons) the increased availability of high performance computing facilities and efficient, easy to use DFT codes. Researchers can choose among several different open source and commercially available \textit{ab-initio} software packages.  The numerical approach used in these packages may vary significantly and may dictate which quantities are more convenient to calculate. For example, the KKR method fits naturally with the coherent potential approximation, yielding efficient total energies for disordered alloy systems \cite{cpa1,cpa2,cpa3}. The Kubo-Greenwood equation in combination with KKR-CPA \cite{butler1985} allows efficient computation of residual resistivity for systems with high chemical disorder, whereas the semi-classical Boltzmann equation may require band structure calculation of a large unit cell \cite{boltztrap}. The modern KKR method solves for the Green's functions of the Kohn-Sham equation rather than the wavefunctions, i.e., the Kohn-Sham orbitals. The optimal choice of software package hence depends on the physical properties that the user intends to calculate. It is important to ensure that the value of a given physical property for a given system is consistent, regardless of the computational method. This consistency provides confidence in the software implementation, and allows a new first-principles code to establish it's validity.

Many parameters like lattice constant, bulk moduli etc could be chosen to compare different codes \cite{fuyangpaper}. Our focus is on the consistency of total energies. The total energy (as a function of concentration) is necessary for the computational study of phase stability and phase transitions. For the order-disorder transition in CuZn, for example, the calculated critical temperature can be very sensitive to the energies used. In order to use any DFT technique to study this transition, it is important to ensure the consistency of the first-principles energies. 

While there are many ways to solve the Kohn-Sham equation numerically, we look at the three major approaches - Full-potential linearized augmented plane Waves (FLAPW)\cite{lapw,flapw,flapw-energy}, plane wave based pseudo-potential \cite{plane-wave-pseudo}, and the KKR Green's Function\cite{korringa,kohnrostoker} techniques as implemented in the software packages WIEN2k \cite{wien2k}, VASP \cite{vasp}, and MuST \cite{must}, respectively. Using these three codes, we calculate the total energy for different arrangements of BCC binaries AB where A,B $\in$ \{Cr, Mo, Nb, V\} and of  FCC binaries \textit{CD} where  C,D $\in$ \{Ag, Cu, Ni, Pd\}. 

The paper is organized as follows. First we summarize the theory behind the three DFT techniques and highlight the similarities, differences, advantages and disadvantages of each approach. We then present our first-principles energy calculations for the chosen structures, analyze their consistency and how it varies among different binary compounds. We also look at the impact of some standard DFT input parameters on the results obtained.

\section{Theory}
Density functional theory codes express the multi-electron Schr\"odinger equation, a complex function of position vectors of each electron, as a single-electron equation, referred to as the Kohn-Sham equation \cite{hohenberg,kohnsham}
\begin{equation}
    \left[-\nabla^{2} + V_{\rm{eff}}(\left[\rho(\bm{r})\right])\right]\psi_{i}(\bm{r}) = \epsilon_{i}\psi_{i}(\bm{r}).
\end{equation}
where the Hamiltonian is a functional of the density $\rho(\bm{r}) = \sum_{\epsilon_i < \epsilon_F} \vert \psi_i(\bm{r}) \vert^2$ and we set $\hbar = 1$ and $m_e = 1/2$. This implicit equation in $\psi$ has to be solved self-consistently. However, as mentioned previously, there are several alternative routes to solving it.

\subsection{Full-potential Linearized Augmented Plane Wave}
The LAPW method involves breaking the system into atomic (or ``muffin-tin") spheres, with each sphere containing a single atom.  The wavefunction in the muffin-tin sphere can be expressed as a product of radial functions and spherical harmonics\cite{wien2k,dsinghbook}
\begin{equation}
    \phi_{\bm{k}_n} = \sum_{lm} \left[A_{lm,\bm{k}_n}u_{l}(r, E_l) + B_{lm,\bm{k}_n}\dot{u}_{l}(r,E_l)\right]Y_{lm}(\hat{\bm{r}}),
    \label{eq:linearized-function}
\end{equation}
where $u_l$ is the regular solution of the radial Schrodinger equation with energy $E_l$ and $\dot{u}_l$ is its energy derivative taken at the same energy. This is a linear expansion of an energy-dependent function about the fixed energy $E_l$ (linearization). The subscript $\bm{k}_n = \bm{k} + \bm{K}_n$, where $\bm{k}$ is the wavevector within the Brillouin zone and $\bm{K}_n$ is a reciprocal lattice vector.  The coefficients $A_{lm},B_{lm}$ are obtained by matching the wavefunction and it's derivative on either side of the muffin-tin boundary. In early applications of the LAPW method, the potential inside the muffin-tin sphere was taken to be spherical (muffin-tin approximation). Full-potential LAPW (FLAPW) removes this restriction. The consequence of assuming spherical symmetry is explored in detail in the next section. 

This form of the wavefunction is capable of describing low energy electrons (like 1\textit{s}), which are highly localized around the nucleus and are not strongly affected by the electrons from the other atoms in the crystal. Such electrons are termed core electrons. For  higher energy electrons, the interstitial region becomes important. In the interstitial region, the wavefunction
\begin{equation}
    \phi_{\bm{k}_n} = \frac{1}{\sqrt{w}}e^{i\bm{k}_n\cdot\bm{r}}
    \label{eq:plane-wave}
\end{equation}
can be expressed as a plane wave expansion. Overall, $\phi_{\bm{k}_n}$ can be considered as a plane wave augmented by a linearized spherically decomposed function inside the muffin-tin sphere, as specified by Equation (\ref{eq:linearized-function}). Hence the method is called Linearized Augmented Plane Wave (LAPW). The overall solution of the system can be written as
\begin{equation}
    \psi_{\bm{k}} = \sum_{n} c_n \phi_{\bm{k}_n}.
\end{equation}
The coefficients $c_n$ are obtained using the Rayleigh-Ritz variational principle. Since this method explicitly deals with the core and the valence electrons, it is termed as an all-electron method. Interested readers may refer to David Singh's book \cite{dsinghbook} to learn more about the FLAPW method.

\subsection{Pseudopotential method}
In this method, the valence electron wavefunction is represented using a basis of plane waves
\begin{equation}
    \psi_{\bm{k}} = \sum_{n} c_{\bm{k}_n}e^{i\bm{k}_n\cdot\bm{r}}.
    \label{eq:pw}
\end{equation}
This is a simpler form of the wavefunction than in the augmented plane wave approach, but it is only useful for the valence electrons. Core electrons, due to their strong localization, require an impractically large number of plane waves to adequately describe the state. To overcome this problem, the ``frozen core" approximation is employed \cite{frozencore}. The core electrons are assumed to contribute a constant offset to the total energy and are hence calculated (and stored) separately. To remove the core electrons from the main calculation, the effective potential is replaced by a pseudopotential that replaces the nuclear $1/r$ behavior with a screened form that remains finite at $r = 0$. The Kohn-Sham equation then becomes
\begin{equation}
    \left[-\nabla^2 + V^{\rm{ps}}_{\rm{eff}}([\rho(\bm{r})])\right]\psi^{\rm{ps}}_{i}(\bm{r}) = \epsilon_i\psi^{\rm{ps}}_{i}(\bm{r}),
\end{equation}
where $V^{\rm{ps}}_{\rm{eff}}$ is the pseudopotential and $\psi^{\rm{ps}}_{i}$ is the pseudo-wavefunction, which is solved by expanding in terms of plane waves as shown in Equation \ref{eq:pw}. Due to the freezing of core electrons, this is not an all-electron method.

There are many types of pseudopotentials, each having it's own advantages and disadvantages. The general goal is to generate ``smooth" pseudopotentials which provide highly accurate results with fewer plane wave basis elements. Popular examples are the Vanderbilt ultrasoft pseudopotentials\cite{ultrasoft} and the projector augmented wave (PAW) potentials\cite{paw},  and we check the latter in our calculations utilizing the plane-wave code VASP \cite{vasp}.

\subsection{KKR Green's Function method}
This technique resembles FLAPW as an all-electron method that divides the system into non-overlapping and space filling atomic cells. However, instead of solving for the Bloch wavefunction, the modern KKR method obtains the spacially resolved Green's function of the Kohn-Sham equation in each atomic cell. This is done using multiple scattering theory: each atomic cell is treated as an electron scattering center, for which the local (or ``single-site") solutions and the scattering $t$-matrix can be calculated. The $t$-matrices can be used to construct the multiple scattering path matrix $\tau$ \cite{Gyorffy_Stott_1973}, where
\begin{equation}
\underline{\tau}^{nm}(\epsilon)=\underline{t}^{n}(\epsilon)\delta_{nm}+\underline{t}^{n}(\epsilon)\sum_{k\neq n}\underline{g}^{nk}(\epsilon)\underline{\tau}^{km}(\epsilon)
\label{eq:tau_exp}
\end{equation}
is defined as the sum of all scattering process of an electron travelling from site $n$ to site $m$, with $g^{nk}(\epsilon)$ representing the free-electron propagator. In the vicinity of atomic site $n$, the Green's function \cite{Faulkner_Stocks_PRB_1980}
\begin{equation}
\begin{split}
G_{\sigma\sigma^\prime}({\bf r}_n,{\bf r}^\prime_n;\epsilon)= &\sum_{LL^\prime}Z^{n}_{L\sigma}({\bf r}_n;\epsilon)\tau^{nn}_{L\sigma L^\prime\sigma^\prime}(\epsilon)Z^{n\ast}_{L\sigma^\prime}({\bf r}_n;\epsilon) \\
&-\sum_{L}Z^{n}_{L\sigma}({\bf r}_n;\epsilon)J^{n\ast}_{L\sigma}({\bf r}_n;\epsilon)\delta_{\sigma\sigma^\prime},\label{eq:MSTG}
\end{split}
\end{equation}
is a matrix in the spin space, where  $Z^n_{L\sigma}({\bf r}_n;\epsilon)$ and  $J^n_{L\sigma}({\bf r}_n;\epsilon)$ are single-site solutions with $L \equiv (l,m)$ and $\sigma$ being the electron spin index. 

Many useful physical properties can be obtained from the Green's function. For example, the electron density of the valence states in the vicinity of atomic cell $n$ can be obtained by integrating the Green function from the bottom of the valence band $\epsilon_B$ to the Fermi energy $\epsilon_F$,
\begin{equation}
\begin{split}
\rho^n({\bf r}_n)=-\frac{1}{\pi}{\rm Im}{\rm Tr}\int_{\epsilon_{\rm B}}^{\epsilon_{\rm F}}\underline{G}({\bf r}_n,{\bf r}_n;\epsilon)d\epsilon.
\end{split}
\label{eq:rho}
\end{equation}
This density is then used to calculate the new effective potential with in the LDA or GGA approximation. Since the Green's function expression (Equation \ref{eq:MSTG}) is formally exact, this electron density yields the exact solution of the the Kohn-Sham equation, rather than an approximate solution based on variational principle. Linearized versions of the KKR-Green's functions like LMTO\cite{lmto1,lmto2,lmto3} or EMTO\cite{emto1,emto2,emto3} are also available, providing efficient computation at the cost of exactness. The electronic density of states can also be obtained from the Green function,
\begin{equation}
		D^n({\epsilon})=-\frac{1}{\pi}{\rm Im}{\rm Tr}\int_{\Omega_n} \underline{G}({\bf r}_n,{\bf r}_n;\epsilon)\;d^3{\bf r}_n.
	\label{eq:De}
\end{equation}
However, the Bloch wavefunctions, i.e., the Kohn-Sham orbitals, and their dispersion relations, or the band structures, are not conveniently given. 

\section{Computational details}
\label{sec:comp-details}
To compare the energetics of differing atomic configurations we created six distinct 16-atom BCC structures (Figure \ref{fig:BCC_unitcells}) and four distinct 8-atom FCC structures (\ref{fig:FCC_unitcells}). A minimum mesh of 14x14x14 K-points were used for the 16-atom BCC structures and 19x19x19 K-points were used for the 8-atom FCC structures to converge the energies. The PBE generalized gradient approximation\cite{pbe} was used. For VASP, we chose the PAW potentials (Cr\_pv 2007, Mo\_pv 2005, Nb\_pv 2002, V\_pv 2000, Ag\_pv 2005, Cu\_pv 2000, Ni\_pv 2000, Pd\_pv 2005). For MuST and WIEN2k, the deep-core states were treated in a fully relativistic manner with spherical potential approximation, and the valence states were treated with full-potential scalar relativity. In the MuST calculations, the semi-core states were treated in the same manner as the deep-core states, while in the WIEN2k calculations, these semi-core states were included in the band structure calculations so that in essence, they were treated in a full-potential scalar relativistic manner. The maximum orbital angular momentum quantum number $l_{\rm{max}} = 4$ was chosen for MuST and $l_{\rm{max}}=10$ for WIEN2k. Later we discuss the impact of certain choices.

\section{Relative Energy Comparison}
\subsection{BCC binaries}
\begin{figure}
	\centering
	\includegraphics[width=0.5\textwidth]{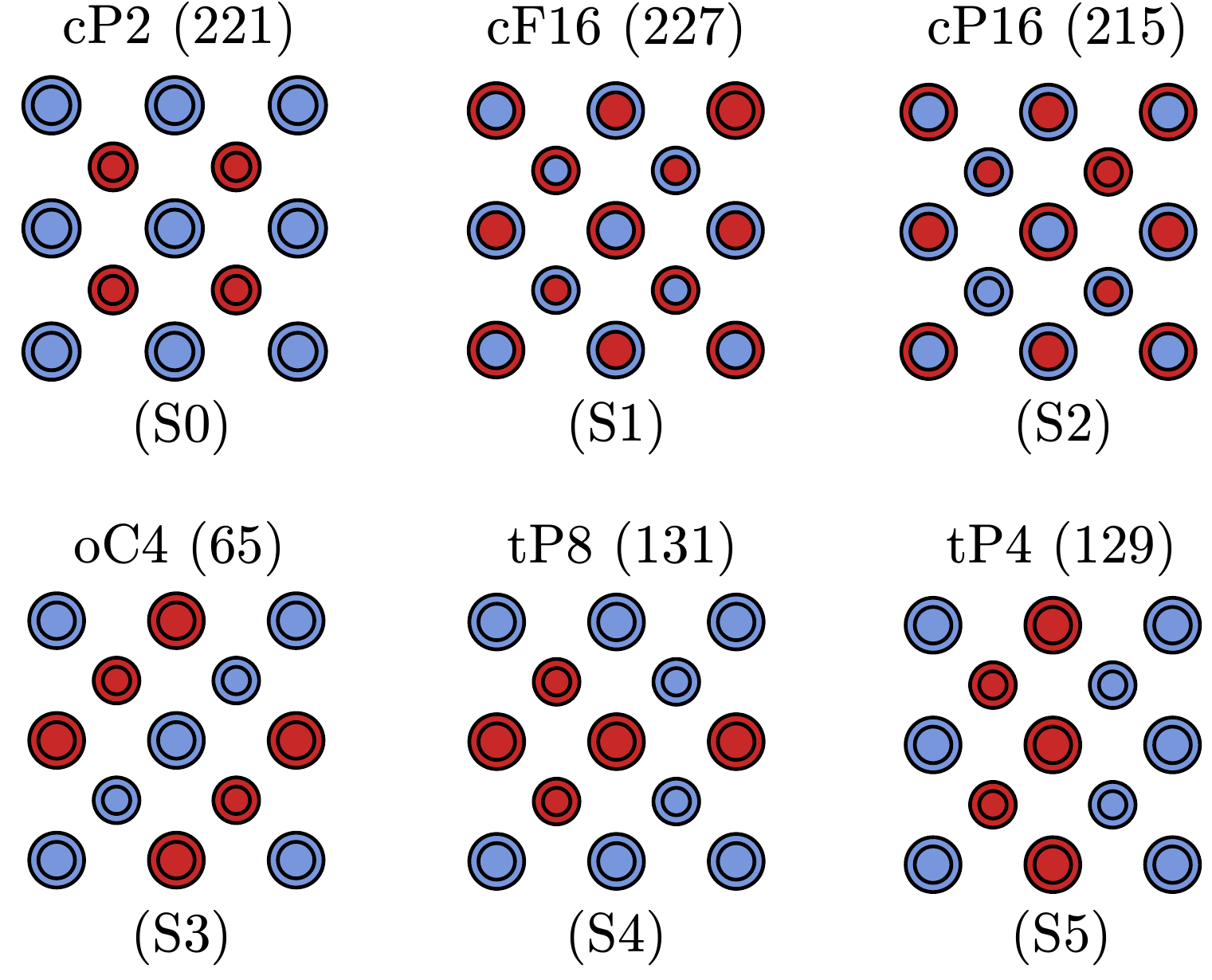}
	\caption{Unit cells of the six BCC structures, labelled using structure indices S0-S5. Above each structure its structure is decribed in the format [Pearson symbol](Space Group). Color indicates chemical species and size indicates vertical height (small on top)}
	\label{fig:BCC_unitcells}
\end{figure}

We focus on the six binary alloys obtained from the BCC refractory elements Cr, V, Mo and Nb. For each binary, we look at a set of six 16-atom structures with differing symmetries, as presented in Figure \ref{fig:BCC_unitcells}. The total energy for all 36 of these structures are calculated using VASP, WIEN2k and MuST. The PAW pseudopotentials have been tuned using an FLAPW method \cite{paw}, which explains the excellent agreement between VASP and WIEN2k relative energies. The MuST calculations have been performed in two modes - a muffin-tin treatment, where the potential and the electron density are assumed to be spherically symmetrical, and a full-potential treatment of the valence electrons.
\begin{figure}
	\includegraphics[width=\linewidth]{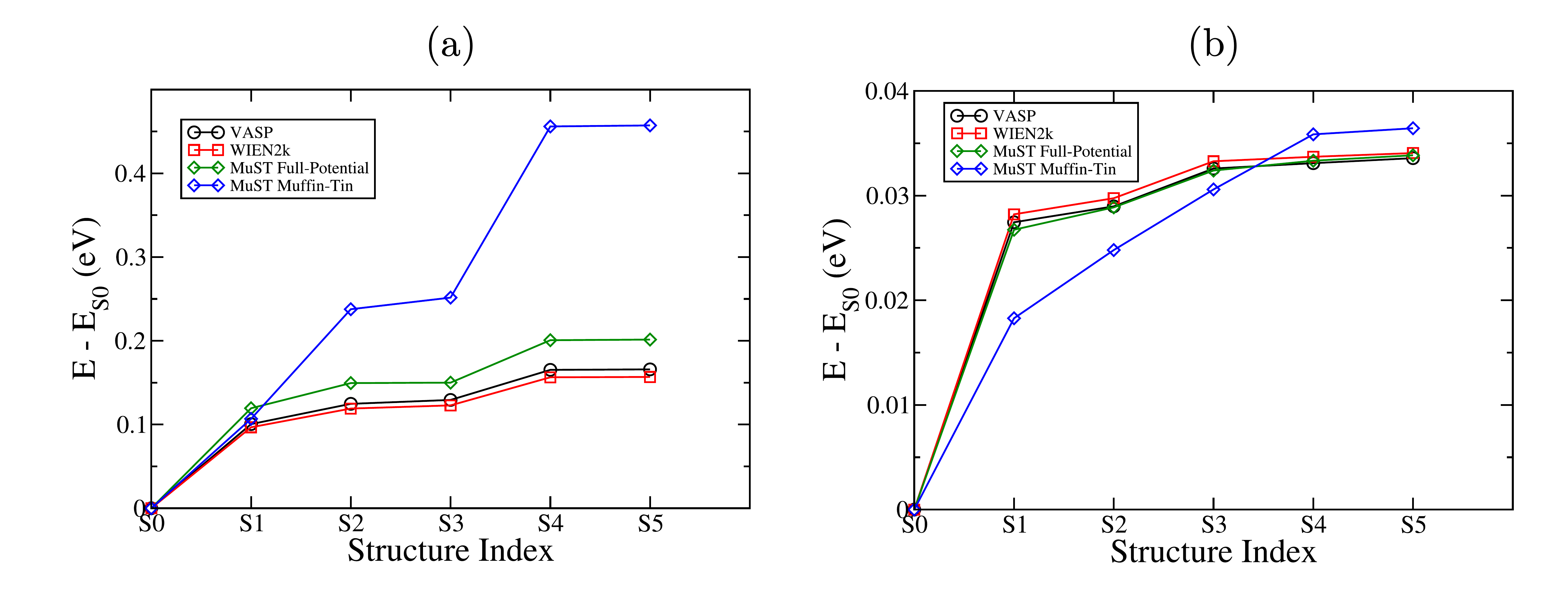}
	\caption{Energies of the structures S1-S5 relative to S0 as calculated using VASP, WIEN2k and MuST (with full-potential and muffin-tin treatment of valence electrons) for (a) CrNb (b) CrV.}
	\label{fig:BCC-rel-energies}
\end{figure}
\begin{figure}
	\includegraphics[width=0.6\linewidth]{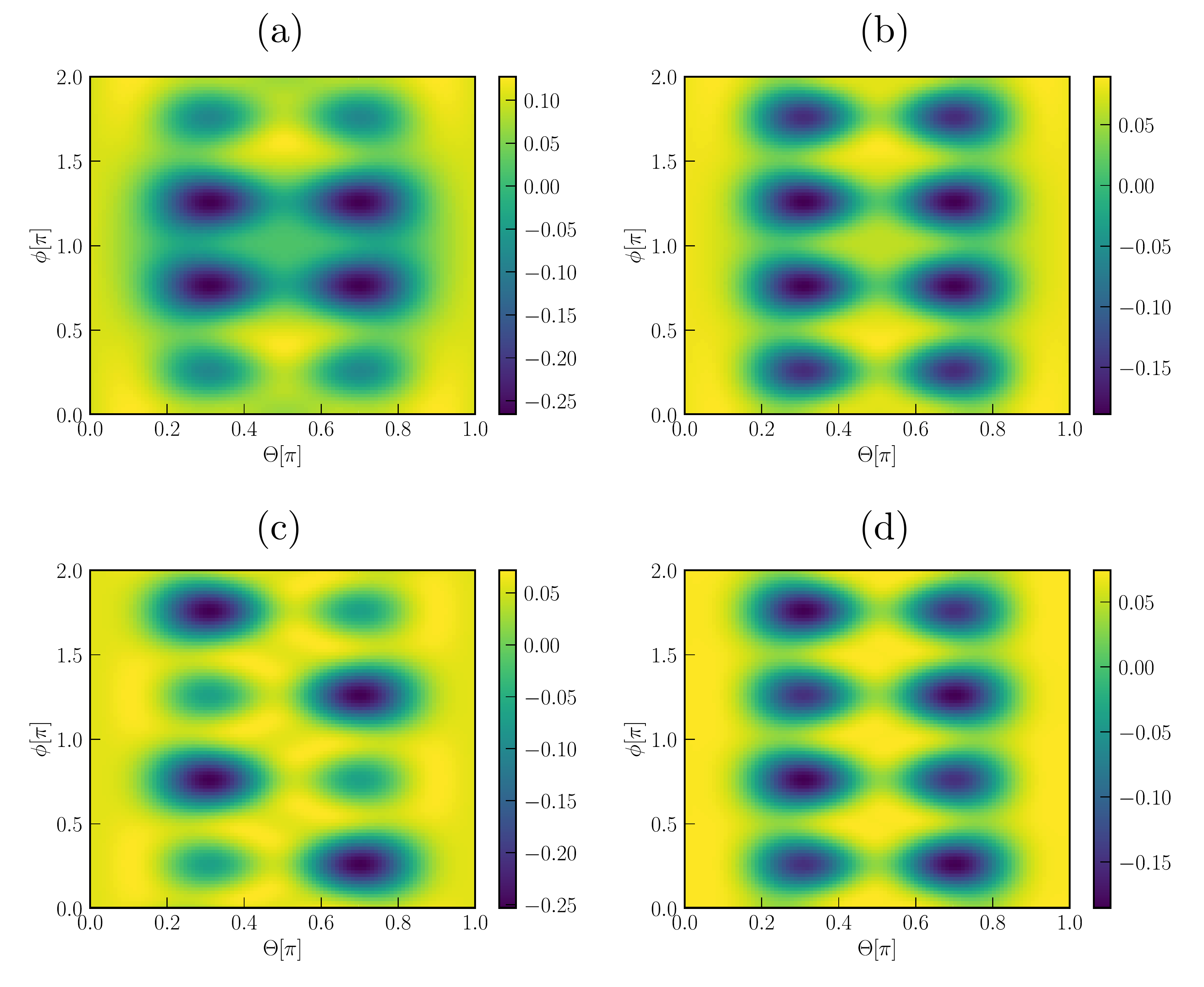}
	\caption{Plotting the angular potential ($\sum_{lm}V_{lm}Y_{lm}$) in  the atomic cell centered at (0,0,0) as a function of angles $\Theta, \phi$ for (a) CrNb S5 (b) CrV S5 (c) CrNb S1 (d) CrV S1 }
	\label{fig:Vplots}
\end{figure}
\begin{table}[]
	\begin{tabular}{|c|c|c|c|c|c|c|c|}
		\hline
		& & S0 & S1 & S2 & S3 & S4 & S5 \\ \hline
		\multirow{2}{*}{CrNb} & Number of $lm$ Components & 8 & 11 & 27 & 15 & 25 & 25  \\ \cline{2-8}
		& $\sum_{lm \neq(0,0)} \vert V_{lm} (r_{\rm{mt}}) \vert^2 $  & 0.0788 & 0.0451 & 0.0605 & 0.0579 & 0.0647 &  0.0641 \\ \hline
		\multirow{2}{*}{CrV}  & Number of $lm$ Components & 8 & 11 & 22 & 13 & 20 & 20  \\ \cline{2-8}
		& $\sum_{lm \neq(0,0)} \vert V_{lm} (r_{\rm{mt}}) \vert^2 $   & 0.0479 & 0.0399 & 0.0442 & 0.0425 & 0.0435 & 0.0435 \\ \hline
	\end{tabular}
	\caption{The number of $(l,m)$ components that contribute significantly to the non-spherical part of the atomic potential; the norms of the angular terms $\sum_{lm \neq(0,0)} \vert V_{lm} (r_{\rm{mt}}) \vert^2$, in units of Ryd$^2$.}
	\label{tab:lm-comp}
\end{table}
\begin{table}[]
	\begin{tabular}{|c|c|c|}
		\hline
		& Time Taken (s) & Approximate Memory Usage   \\ 
		&  &   (MB per MPI process) \\ \hline
		VASP                                & 816            & 353                                                  \\ \hline
		WIEN2k (full unit cell)             & 7387           & 117                                                   \\ \hline
		MuST Muffin-Tin                           & 33054          & 335                                                  \\ \hline
		MuST Full Potential                             & 57694          & 543                                                  \\ \hline
	\end{tabular}
	\caption{Performance analysis of the different DFT codes applied to the CrV S0 structure on a 16-core node. Time taken is measured from execution to SCF convergence, with similar number of SCF iterations in all cases. Here memory usage refers to the resident memory, \textit{{i.e}} how much RAM the process is actively using.}
	\label{tab:performance}
\end{table}

Out of the six compositions, the energies for CrNb and CrV, are shown in Figure  \ref{fig:BCC-rel-energies}. The energies are plotted relative to those of the S0 structure. For CrNb, the muffin-tin relative energies show large deviations, while for CrV the muffin-tin energies are much closer to the corresponding WIEN2k values. This implies that the muffin-tin treatment of valence states is not a good approximation for CrNb, but works well for CrV.  

From the remaining four binary compounds , CrMo, MoV and NbV have similar relative energy behavior as CrNb, while MoNb, like CrV, shows small relative energy difference between WIEN2k and MuST, even with the muffin-tin approximation. It seems that atomic potentials vary more strongly between rows of the periodic table than between adjacent columns. 

To understand this, write the potential in an atomic cell $n$ of the structure $S$ 
\begin{equation}
	V^n(\bm{r}_n) = \frac{1}{r}V^n_r(r_n) + \sum_{lm \neq (0,0)}V^n_{lm}(r_n)Y_{lm}(\hat{\bm{r}_n}),
\end{equation}
where $\bm{r}_n$ is the radial vector from the center of the atomic cell, $V_r$ represents the spherical part, and the $V_{lm}$ coefficients represent the deviation from spherical symmetry.  Figure \ref{fig:Vplots} plots the second term of the potential at the muffin-tin radius for the atomic cell centered at the origin of the S1 (cF16) and S5 (tP4) structures of CrNb and CrV (Figure \ref{fig:Vplots}). In all plots, there are peaks at $\theta,\phi = (2n + 1)\pi/4$, corresponding to the positions of near-neighbor atoms in a BCC structure. For the CrNb structures the strengths of the peaks vary in accordance with the chemical ordering of the structure. This variation is not seen for the CrV structures and the peaks seem similar in strength. This implies that the CrNb potential is more anisotropic and as a result is poorly described by the muffin-tin approximation.

To quantify the extent of asymmetry in the potential, we count how many terms in the $lm$ sum contribute to the potential. If, for any atomic cell $i$, and any radius $r_i$, $V^i_{lm}(r_i) > 10^{-4}$ Ryd, we consider the corresponding $lm$ index pair to contribute to the asymmetry of structure $S$. We also calculate $\sum_{lm \neq 0} \vert V_{lm}(r_{\mathrm{mt}})\vert^2$, which quantifies the overall asymmetry in the system, summed over all the angular modes. Table \ref{tab:lm-comp} shows that both these factors are lower for CrV, implying weaker angular dependence in the system. This could help explain the relative energy results seen in Figure \ref{fig:BCC-rel-energies}.

For the CrNb binaries, despite full-potential valence, there is a maximum energy difference of approximately 45 meV between MuST and WIEN2k at structure S5.  This is largely due to the treatment of semicore states, specifically 3{\it s} and 3{\it p} states for Cr and V, and 4{\it s} and 4{\it p} for Nb, in the two codes. In WIEN2k, the semicore states are treated as valence with a full-potential representation, whereas in MuST the semicore states are considered as local bound states, and are hence treated with the muffin-tin approximation. Unfortunately we were unable to converge the MuST calculation when treating semicore states as valence. The differing semicore treatment does not seem to have an effect on the CrV structures, for the above-mentioned reasons. 

A performance analysis of the three programs was performed by applying them to the CrV S0 structure and tracking the time and total memory used. The results are presented in Table \ref{tab:performance}. Note that the MuST code has been less optimized than the other two and performance improvements are in progress. The FP MuST calculation is significantly more time-consuming than the muffin-tin, which is good incentive to reduce the energy mismatch between the codes while maintaining the muffin-tin approximation. This will be explored in the next section.

\subsection{FCC binaries}
\begin{figure}
	\centering
	\includegraphics[width=\textwidth]{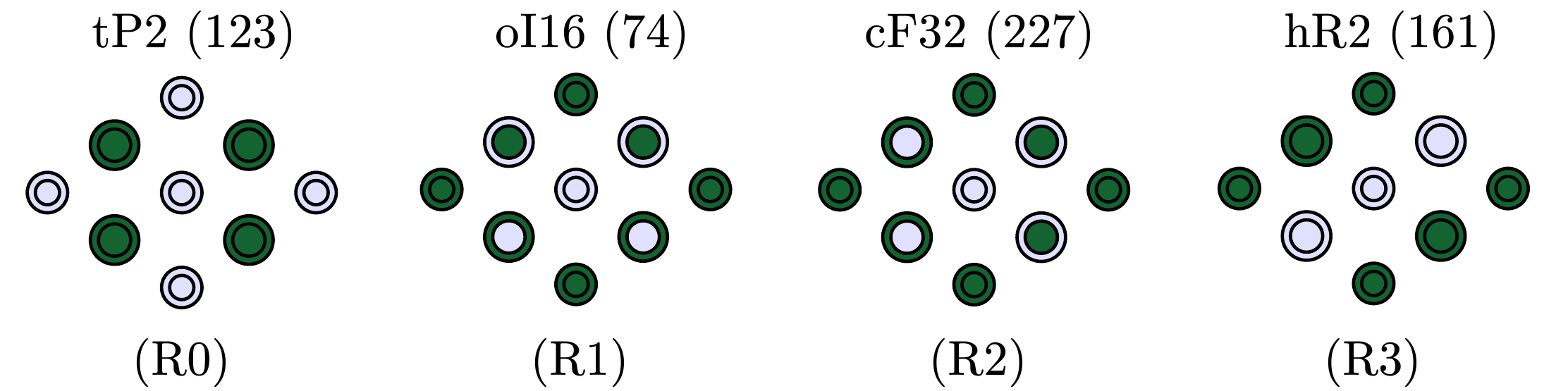}
	\caption{Unit cells of the four FCC structures, labelled using structure indices R0-R3. Above each structure it's symmetry is mentioned, in the format [Pearson symbol](Space Group). Color indicates chemical species and size indicates vertical height.}
	\label{fig:FCC_unitcells}
\end{figure}
\begin{figure}
	\includegraphics[width=\linewidth]{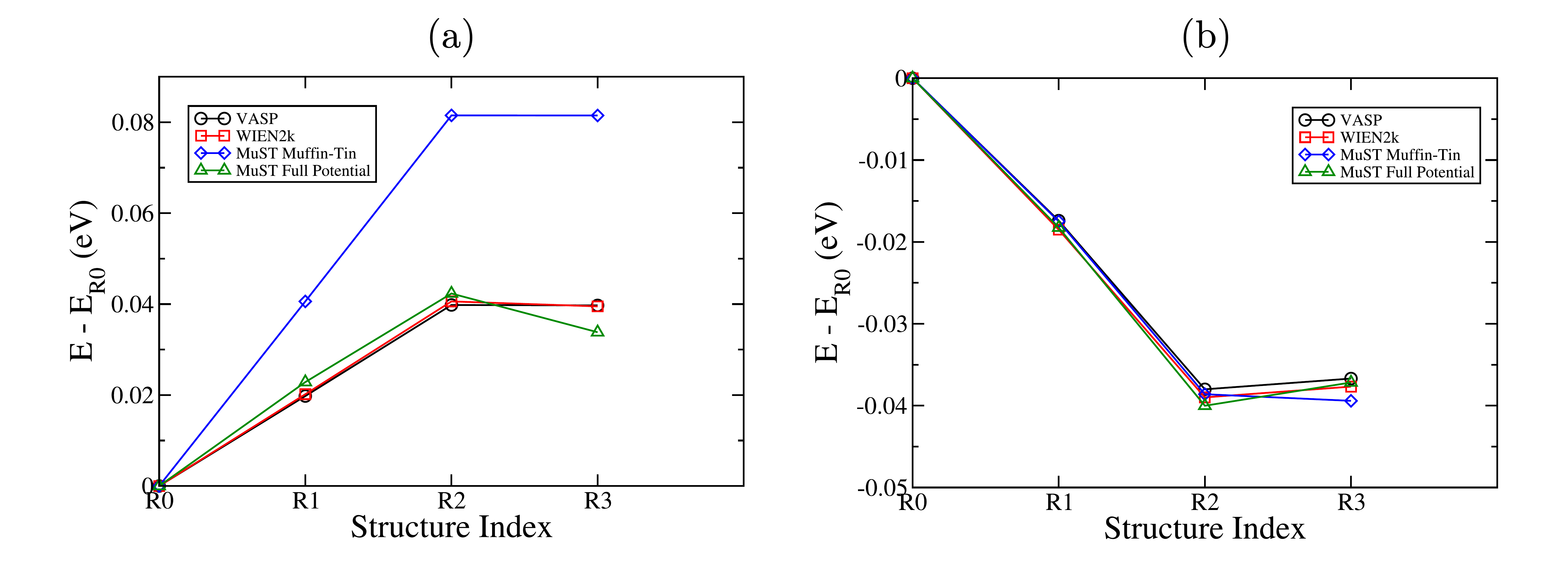}
	\caption{Energies of the structures R1-R3 relative to R0 as calculated using VASP, WIEN2k and MuST (with full-potential and muffin-tin treatment of valence electrons) for (a) NiPd (b) CuNi.}
	\label{fig:FCC-rel-energies}
\end{figure}
\begin{figure}
	\includegraphics[width=0.6\linewidth]{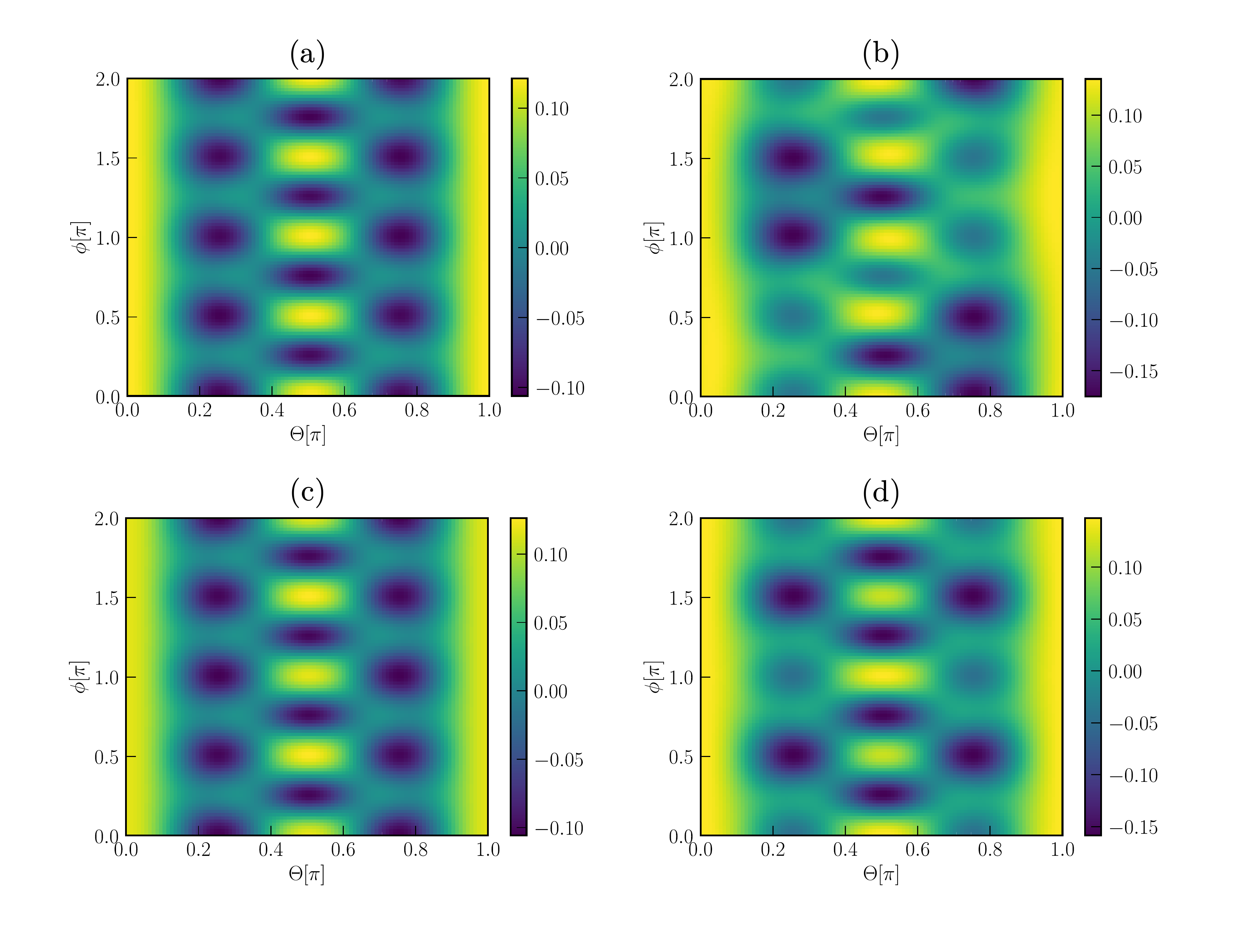}
	\caption{Plotting the angular potential ($\sum_{lm}V_{lm}Y_{lm}$) in  the atomic cell centered at (0,0,0) as a function of angles $\Theta, \phi$ for (a) CuNi R2 (b) NiPd R2 (c) CuNi R0 (d) NiPd R0}
	\label{fig:FCCVplots}
\end{figure}

For the FCC case, we focus on the six binaries obtained from Ag, Cu, Ni and Pd. For each binary composition, we look at four possible 8-atom FCC structures, shown in Figure \ref{fig:FCC_unitcells}. Total energies are calculated for the 24 structures using VASP, WIEN2k and MuST. The behavior is similar to that of the BCC binary alloys. Figure \ref{fig:FCC-rel-energies} shows the relative energies for NiPd and CuNi. The muffin-tin approximation works well for CuNi but badly for NiPd. The reason behind this is similar to the BCC case and can be seen in the potential plots (Figure \ref{fig:FCCVplots}) . The NiPd potentials are more anisotropic, which gives some intuition on why the muffin-tin approximation performs poorly for that system.  Table \ref{tab:FCClm-comp} shows the numbers of $lm$-components that contribute to the non-spherical part of the potential and the sum over the angular-momentum resolved potentials $\sum_{lm \neq 0} \vert V_{lm}(r_{\mathrm{mt}}) \vert^2$. It can be seen that for NiPd, there is a large variation in the number of $lm$-components from R0 to R3, while for CuNi the variation is much smaller. For all structures, the sum over the angular momentum resolved potentials is lower for CuNi, implying a lower asymmetry. From the remaining four binary compounds,  AgPd shows CuNi-like behavior and the others show NiPd-like behavior. 
\begin{table}[]
	\begin{tabular}{|c|c|c|c|c|c|}
		\hline
		& & R0 & R1 & R2 & R3 \\ \hline
		\multirow{2}{*}{NiPd} & Number of $lm$ components & 15 & 25 & 24 & 24  \\ \cline{2-6}
		            & $\sum_{lm\neq(0,0)} \vert V_{lm}(r_{\rm{mt}}) \vert^2 $ & 0.0373 & 0.0354 & 0.0387 & 0.0388\\ \hline
		\multirow{2}{*}{CuNi} & Number of $lm$ components & 19 & 21 & 20 & 18   \\ \cline{2-6}
                    & $\sum_{lm\neq(0,0)} \vert V_{lm}(r_{\rm{mt}}) \vert^2$  & 0.0249 & 0.0253 & 0.0254 & 0.0253 \\ \hline
	\end{tabular}
	\caption{The number of $(l,m)$ components that contribute significantly to the non-spherical part of the atomic potential; the norms of the angular terms $\sum_{lm \neq(0,0)}\vert V_{lm} (r_{\rm{mt}}) \vert^2$ in units of Ryd$^2$.}
	\label{tab:FCClm-comp}
\end{table}
\section{Variation in Muffin-Tin parameters}
We have established that the relative energies vary among different DFT techniques. This difference is especially pronounced comparing the muffin-tin approximation with computationally expensive full-potential methods. It is therefore useful to try and improve the muffin-tin energies. Additionally, in any first-principles study, it is important to ensure that the results obtained are physical and not an artifact of the chosen computational parameters. In this section, we look at BCC CrNb and FCC NiPd (systems for which muffin-tin performed poorly) and identify some computational parameters that may impact the relative energies.
\subsection{Angular Momentum Cutoff}
\begin{figure}
	\includegraphics[width=\linewidth]{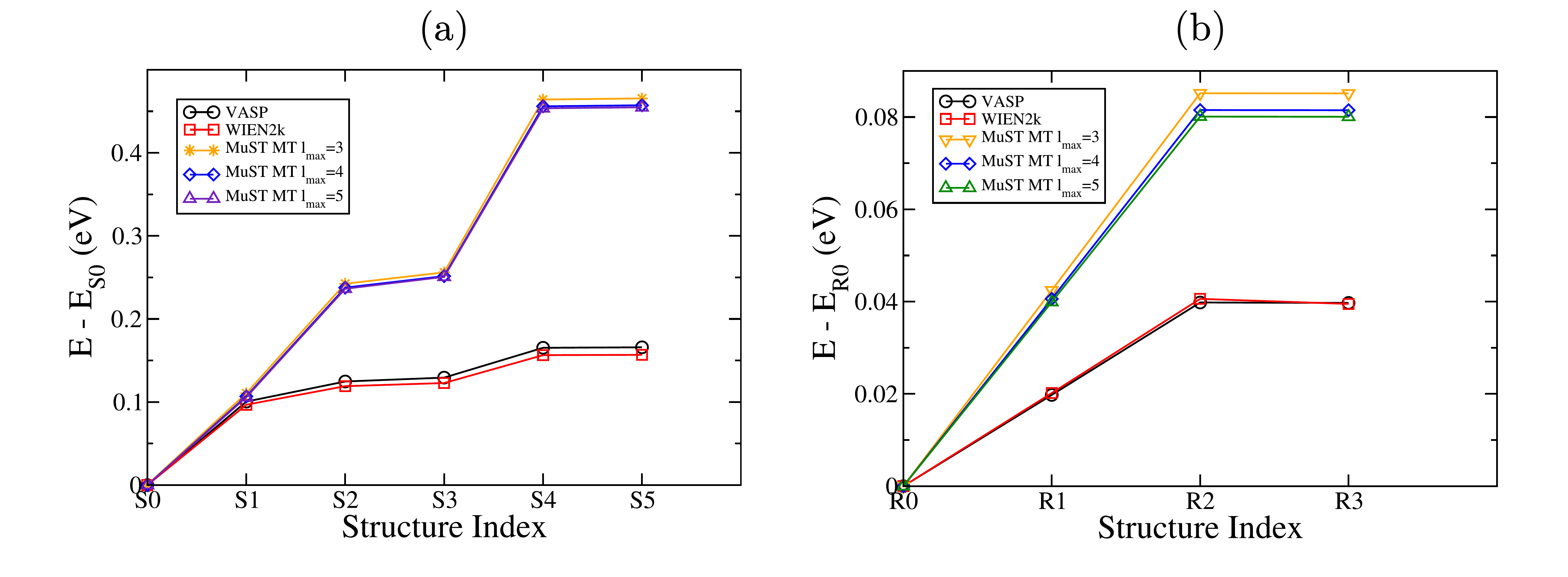}
	\caption{Relative energies calculated using VASP, WIEN2k and muffin-tin MuST with $3 <= l_{\rm{max}} <=5$ for (a) CrNb (b) NiPd}
	\label{fig:lmax-comparison}
\end{figure}

In the KKR method, quantities like the $t$-matrix and $\tau$-matrix are expressed using a spherical harmonic basis. While this is an infinite basis, it has to be truncated for practical use. The angular momentum cutoff (or $l_{\rm{max}}$) determines the size of the truncated basis. Larger $l_{\rm{max}}$ will produce more accurate energies but will incur higher computational cost (scales as $l_{\rm{max}}^6$) . In Figure \ref{fig:lmax-comparison} we show how the relative energies for NiPd and CuNi vary with $l_{\rm{max}}$. It is seen that changing $l_{\rm{max}}$ changes the relative energies, but not enough to bridge the gap between the muffin-tin KKR and the other two methods. 
\subsection{Exchange-Correlation Functional}
All the calculations presented in previous sections used the PBE exchange-correlation functional. We recalculated the VASP, WIEN2k and MuST muffin-tin ($l_{\rm{max}} = 4$) relative energies for CuNb and NiPd using the LDA functional \cite{hohenberg} and obtained results nearly identical to those obtained with PBE.
\subsection{Relativity}
\begin{figure}
	\includegraphics[width=\linewidth]{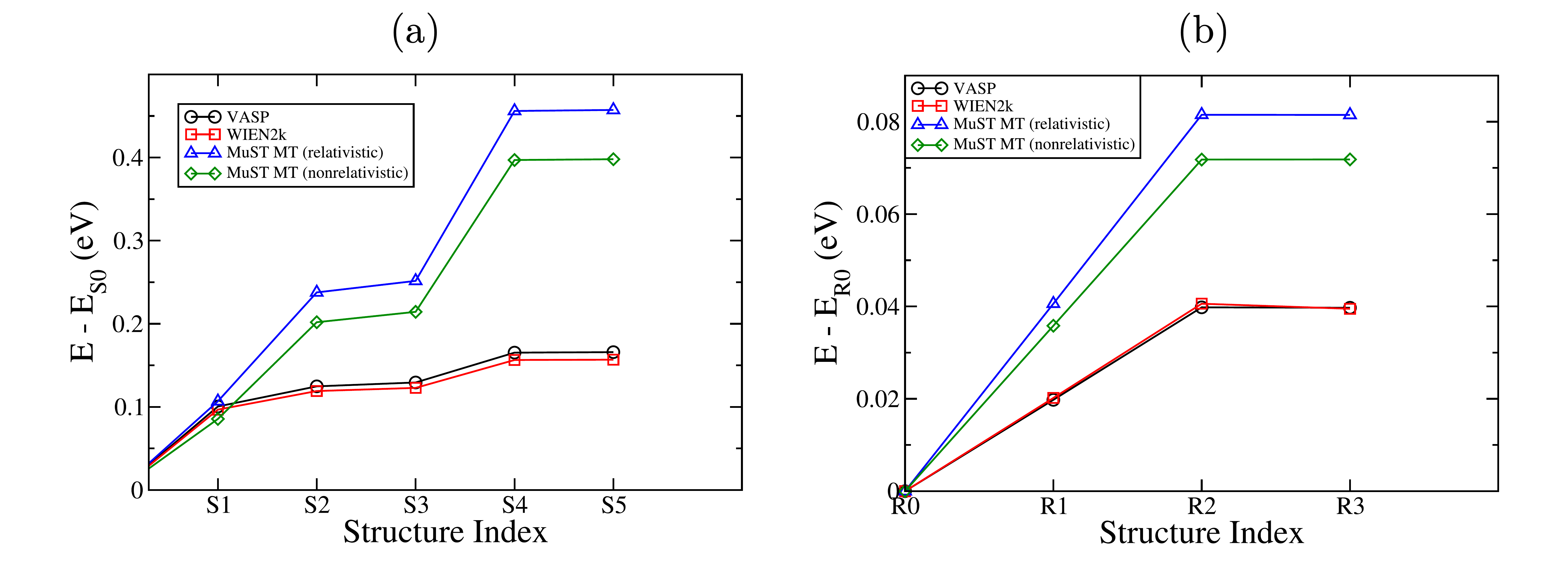}
	\caption{Relative energies calculated using VASP, WIEN2k , relativistic and non-relativistic muffin-tin MuST for (a) CrNb (b) NiPd}
	\label{fig:rel-comparison}
\end{figure}

All the MuST and WIEN2k calculations presented used full relativity for the core and scalar relativity for the valence electrons. To see the effect of relativity, we perform non-relativistic muffin-tin calculations for CrNb and NiPd (shown in Figure \ref{fig:rel-comparison}). The relativistic treatment increases the muffin-tin relative energies and widens the energy gap between MuST muffin-tin and the other methods. Even though the nonrelativistic treatment seems to lower the energy mismatch, it is important to include the relativity in MuST in order to make a fair comparison with WIEN2k and VASP.

\subsection{Radical Plane Ratio}
\begin{figure}
 	\includegraphics[width=\linewidth]{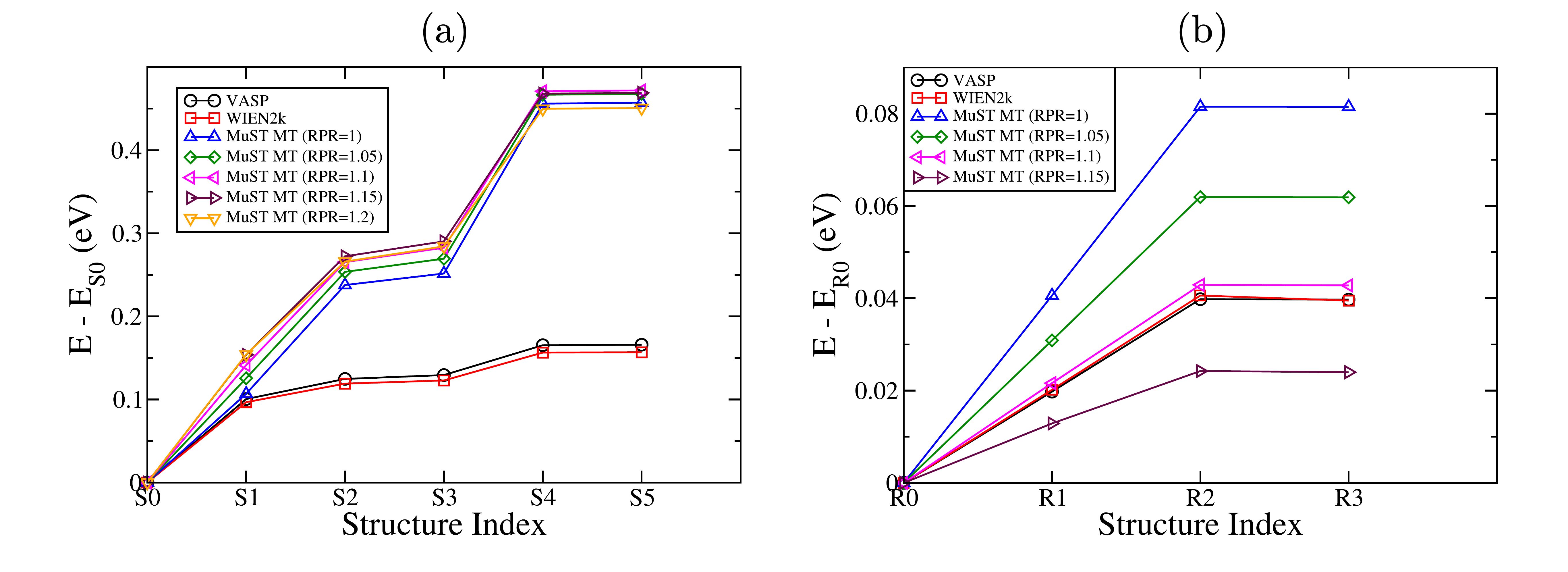}
 	\caption{Relative energies calculated using VASP, WIEN2k and muffin-tin MuST at various RPR values for (a) CrNb (atomic radii ratio 1.192) (b) NiPd (atomic radii ratio 1.134)}
 	\label{fig:RPR-comparison}
  \end{figure}

The muffin-tin approximation involves constructing a sphere centered at each atom, inside which a spherical potential is assumed. The muffin-tin spheres for any two neighboring atoms can touch but do not overlap. The plane that is tangent to two neighboring spheres is referred to as the radical plane. By default, in MuST, the radical plane lies at the center of a line connecting the two neighboring atoms. It may be more preferable for the muffin-tin radii to be consistent with the atomic radii, placing the radical plane closer to the smaller atom. This positioning of the radical plane can be controlled by specifying a radical plane ratio (RPR), which determines the distances from the two atom centers and radical plane, as an input parameter. Figure \ref{fig:RPR-comparison} shows how the relative energies change with RPR. For BCC CrNb, the RPR does not resolve the mismatch between MuST MT and the others. But for FCC NiPd, increasing the RPR moves the muffin-tin relative energies closer to the WIEN2k/VASP values, and at an RPR of 1.1, the relative energy differences between the methods almost vanishes. This shows that for some systems,  the RPR can have a significant impact on the relative energies. 
\section{Conclusion}
Energies obtained from first-principles calculations are sensitive to the underlying approximations employed in  DFT codes. Incorrect energies could produce incorrect elastic constants, phonon behavior, phase transition temperatures, and other important physical quantities that are based on the post-processing of DFT energies.  This paper explores three popular implementations of DFT: pseudopotential plane wave (VASP), full potential linearized augmented plane wave (WIEN2k), and KKR Green's Function method (MuST). The methods are applied to a set of BCC and FCC test systems and the resulting energies are analyzed. The impact of some standard computational parameters on the energies has been explored with a focus on improving the accuracy of the KKR method as implemented by MuST.

Six BCC binary alloys were considered, and for each binary, six distinct configurations were compared. For neighboring elements within a row of the periodic table, the muffin-tin energies compared well with the other methods, while between rows there were large mismatches between muffin-tin and the VASP/WIEN2k/full-potential energies. This mismatch seems to be a consequence of the anisotropy in the potentials, which the muffin-tin approximation cannot handle as it assumes spherical potential. Similar effects were found for binary FCC structures.

The impact of some essential computational parameters like angular momentum cutoff, exchange-correlation potential and radical plane ratio on the muffin-tin energies was also explored. While some small effects of changing $l_{\rm{max}}$ were observed, the mismatch between muffin-tin and the other methods remained. The radical plane ratio however, had a significant impact on the FCC NiPd system and is an important parameter that must be taken into consideration when performing calculations that involve muffin-tin spheres.

\begin{acknowledgments}
This work was supported by NSF under grant DMR-2103958. The authors would like to thank the members of the MuST Program for Disordered Materials, especially Markus Eisenbach and G. Malcolm Stocks for providing valuable feedback.
\end{acknowledgments}

\end{document}